\pdfoutput=1
\documentclass{article}

\usepackage{microtype}
\usepackage{graphicx}
\usepackage{subfigure}
\usepackage{booktabs} 

\usepackage[final]{neurips_2022_ml4ps}
\setcitestyle{square,numbers}

\usepackage[T1]{fontenc}    
\usepackage{hyperref}       
\usepackage{url}            
\usepackage{booktabs}       
\usepackage{amsfonts}       
\usepackage{nicefrac}       
\usepackage{microtype}      
\usepackage{xcolor}         
\usepackage{multirow}
\usepackage{graphicx}
\usepackage{amsmath}
\usepackage{fontawesome}

\title{Neural Inference of Gaussian Processes for Time Series Data of Quasars}

\author{%
  Egor Danilov \\
 Institute of Physics, Laboratory of Astrophysics,\\
École Polytechnique Fédérale de Lausanne (EPFL)\\
  \texttt{egorssed@gmail.com} \\
   \And
  Aleksandra \'Ciprijanovi\'c \\
  Fermi National Accelerator Laboratory\\
  \texttt{aleksand@fnal.gov} \\
   \And
   Brian Nord \\
  Fermi National Accelerator Laboratory;\\
 Kavli Institute for Cosmological Physics \&\\ 
Department of Astronomy and Astrophysics,\\
The University of Chicago; \\
Laboratory for Nuclear Physics, MIT \\
   \texttt{nord@fnal.gov} \\
}

\begin{document}

\bibliographystyle{unsrtnat}

\maketitle

\begin{abstract}

The study of quasar light curves poses two problems: inference of the power spectrum and interpolation of an irregularly sampled time series. 
A baseline approach to these tasks is to interpolate a time series with a Damped Random Walk (DRW) model, in which the spectrum is inferred using Maximum Likelihood Estimation (MLE). 
However, the DRW model does not describe the smoothness of the time series, and MLE faces many problems in terms of optimization and numerical precision. 
In this work, we introduce a new stochastic model that we call \textit{Convolved Damped Random Walk} (CDRW). This model introduces a concept of smoothness to a DRW, which enables it to describe quasar spectra completely.
We also introduce a new method of inference of Gaussian process parameters, which we call \textit{Neural Inference}. 
This method uses the powers of state-of-the-art neural networks to improve the conventional MLE inference technique. 
In our experiments, the Neural Inference method results in significant improvement over the baseline MLE (RMSE: $0.318 \rightarrow 0.205$, $0.464 \rightarrow 0.444$). 
Moreover, the combination of both the CDRW model and  Neural Inference significantly outperforms the baseline DRW and MLE in interpolating a typical quasar light curve ($\chi^2$: $0.333 \rightarrow 0.998$, $2.695 \rightarrow 0.981$). 
The code is published on GitHub \href{https://github.com/deepskies/Neural_GP}{\faGithub}.
\end{abstract}

\section{Introduction}
\label{Sect_intro}
Quasars are astronomical sources powered by the accretion of matter falling onto the black hole in a galaxy's center. 
The accretion onto the black hole is a turbulent process, so the brightness varies intrinsically over time. 
This variability not only allows for studying the region around the black hole \cite{Reverb_mapping} but also enables one to trace the evolution of the Universe through strong gravitational lensing \cite{Time_delay_cosmo}. 
The former requires an interpretable model of radiation variability, whereas the latter requires interpolation of a time series with irregular time sampling. 

Due to the accretion turbulence, the variability of quasar radiation is usually modelled by a stochastic Gaussian process called ``Damped Random Walk'' \cite[DRW;][]{DRW}. 
This model has two main problems: the first problem is that the DRW can model the quasar spectrum on time scales of years but not on the scales of months, where the quasar spectrum steepens. \newline \citep[]{Kelly_2009, Kozlowski_2009, MacLeod_2010, Zu_2013}. 
The second problem is associated with the maximum likelihood estimation (MLE), which is used to infer the parameters of a DRW: this method has convergence issues, high computation complexity, and, sometimes, poor $2\sigma$ constraints on the inferred parameters.

In this work, we aim to solve these two problems of conventional DRW modelling of quasar light curves. 
In Section \ref{Section_CDRW}, we introduce a new physically-interpretable model \textit{Convolved Damped Random Walk} (CDRW) that well describes the quasar spectrum on time scales, from days to decades. 
In Section \ref{Section_Neural_inference}, we propose the \textit{Neural Inference} method that aims to solve problems of maximum likelihood estimation. 
Our approach is task-specific in contrast to the generalized approach in \cite{Task_agnostic_Neural_GP}.
In Section \ref{Sect_data}, we present the data.
In Section \ref{Sect_results}, we present the experiments. 
We conclude in Section \ref{Sect_conclusion}.  

\section{Methodology}
\subsection{Convolved Damped Random Walk model of quasar radiation}
\label{Section_CDRW}

\begin{figure}[!t]
   \begin{minipage}[t]{0.48\textwidth}
     \centering
     \includegraphics[scale = 0.375]{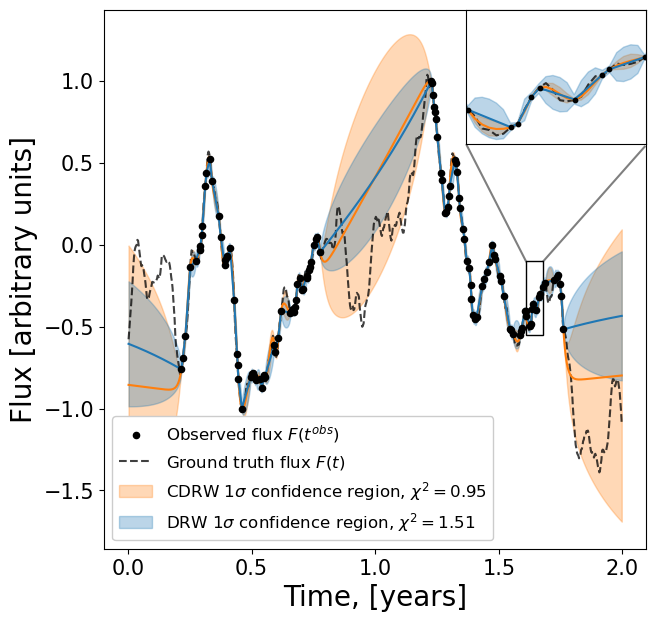}
     \caption{Comparison of Convolved DRW and DRW models in the interpolation of a typical mock quasar light. The envelopes show $1\sigma$ confidence regions. CDRW has better mean and uncertainty of interpolation both in observed regions (upper right corner) and gap regions (middle).
     }\label{Figure_interpolation}
   \end{minipage}\hfill
   \begin{minipage}[t]{0.48\textwidth}
    \centering
    \includegraphics[scale = 0.375]{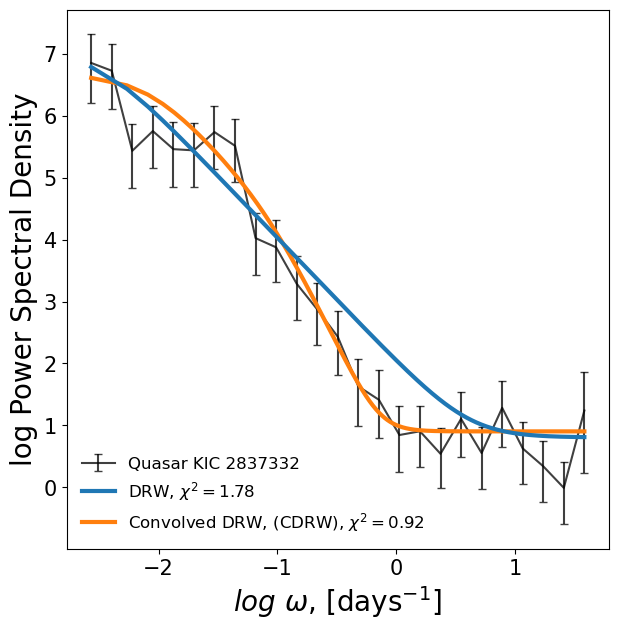}
    \caption{Power Spectrum of a real quasar \citep[black; ][]{Kepler}, baseline DRW \citep[blue $\rho=0$; ][]{Ivezic_Quasar_optical_DRW} and CDRW (orange) introduced in this paper. 
    Error bars and $\chi^2$ are derived from MCMC sampling \cite{Uttley,Timmer_Koenig}. 
    CDRW better describes the power spectrum at high frequencies.}\label{Figure_spectra}
   \end{minipage}
\end{figure}

The conventional physical model of quasar light curve $F(t)$ is given by the lamp-post radiation $g_{\mathit{LP}}$ and thin-disk reverberation $P_{\mathit{TD}}$:

\begin{equation}
    F(t) \propto \int_0^{\infty}     g_{\mathit{LP}}(t-r,\tau)P_{\mathit{TD}}(r,\rho)dr,
    \label{Eq_Quasar_flux}
\end{equation}
where $\tau$ and $\rho$ are the characteristic times of radiation from accretion onto the black hole and from its re-radiation by the outer disk, respectively \cite{Sunyaev_QSO_theory}. 
Usually, $F(t)$ is modelled as a DRW, but this model fails to describe the power spectrum at high frequencies (Fig. \ref{Figure_spectra}). 
This discrepancy can be mitigated if the driving lamp-post radiation $g_{LP}(t)$ is assumed to be DRW instead. 
In this case, convolution with thin-disk reverberation kernel $P_{TD}$ suppresses the power spectrum at high frequencies. 
The spectrum of this kernel is well fitted by a softened powerlaw. 
Such a  derivation brings us to a new model of quasar radiation that we call \textit{Convolved Damped Random Walk} (CDRW) with power spectrum:
\begin{equation}
    PSD_{CDRW}(\omega) =     \frac{\sigma^2}{(1+\omega^2\tau^2)(1+\omega^2\rho^2)},
    \label{eq_CDRW_spectrum}
\end{equation}

where $\omega$ is the circular frequency $2\pi f$. 
Given $\tau \geq \rho$, $\tau$ is correlation time. 
In turn, the reverberation time $\rho$ defines the smoothness of the curve. 
This model well describes the power spectra of real quasars on all frequencies (Fig. \ref{Figure_spectra}).
The newly introduced smoothness of $\rho$ results in better interpolation performance of the model (Fig. \ref{Figure_interpolation}). 
Moreover, in border cases, it reduces to the main types of stochastic processes: white noise, Matern $\nu=1.5$ noise, DRW, and Random Walk (red noise).

\subsection{Neural inference of Gaussian processes}
\label{Section_Neural_inference}
Without prior knowledge of the quasar flux $F$ observed at times $t^{obs}$, the baseline approach to the inference of a Gaussian process is MLE, that is, maximization of likelihood $p(F(t^{obs}) | \mu,\sigma^2,\tau,\rho)$. 
In the context of numerical methods, MLE requires matrix inversion that has computational complexity $O(N^3)$. 
For light curves with a high number of observations $N$, such an inference requires much computational power, moreover, the numerical errors in matrix inversion can jeopardize the inference results. 
In the context of the theory of optimization, MLE faces convergence issues due to orders of magnitude differences between gradients on different parameters. 
Additionally, there are degeneracies between the parameters and biases of modes of the posterior distribution -- e.g. Bessel correction for inverse Gamma posterior of sample variance \cite{Inverse_Gamma_distr}.

To mitigate these negative aspects of Gaussian process inference, we introduce a novel method that we call \textit{Neural Inference}. We use a neural network to predict the final correlation time $\tau$ and an initial guess of reverberation time $\rho$. 
In this case, maximization of the likelihood $p(F(t^{obs}),\tau | \mu, \sigma^2, \rho)$ leads to significantly more accurate parameters. 
Moreover, these estimations become more stable with respect to the stochastic realisation of the studied Gaussian process.
\begin{figure}
    \centering
    \includegraphics[scale=0.55]{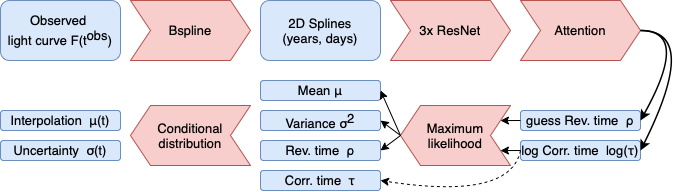}
    \caption{Scheme of light curve interpolation based on the introduced Neural inference method.}
    \label{Fig_Scheme_interpolation_method}
\end{figure}
\section{Data}
\label{Sect_data}
We carry out experiments on synthetic data that mimic real observations. 
The observations of quasars have $135 \pm 25$ days-long annual gaps due to occultation by the Sun and bad weather. 
Within the observation season in the $n$-th year, we model time sampling as a stochastic process:

\begin{equation}
    t^{obs}_{n} = \{t_{ni} \ | \  t_{ni} = t_{n0} + i + PoissonP_i(A=0.34,\frac{1}{\lambda} = 4) + \epsilon_{i} \  [days], \epsilon_i \sim \mathcal{U}(\frac{-5}{24},\frac{5}{24})\},
    \label{eq_time_sampling}
\end{equation}
where $PoissonP$ is a discrete Poisson process \cite{Poisson_process_Partition}, that models short observation gaps, and $\epsilon$ is time in the night.
The statistical form and parameters of the observation schedule correspond to the COSMOGRAIL dataset \cite{COSMOGRAIL}. 

In this research, we consider 10 years-long light curves, based on the planned duration of the LSST survey \cite{LSST}. 
For training the neural network, we simulate the mock quasar radiation as a CDRW process (eq. \ref{eq_CDRW_spectrum}):

\begin{equation}
    F(t^{obs}) \sim \mathcal{N}(\mu, \Sigma_{CDRW}(t^{obs},\sigma^2,\tau,\rho)), \quad log(\tau) \sim \mathcal{U}(1.05,2.76), \quad \rho \sim \mathcal{U}(0.55,10).
\label{eq_dataset_priors}
\end{equation}

The distributions were derived from the previous observations of quasars -- including both short-cadence observations and long-duration surveys \cite{Kelly_2009,Yu_2020}.
Specifically, $\rho_{min}=0.55$ reduces CDRW to DRW, which for $\tau_{max} \approx 575$ reduces to RW. 
The value $\rho_{max}=10$ results in a break of the spectrum at the scale of two months, which outlines the spectral breaks reported in the papers \citep[]{Kelly_2009, Kozlowski_2009, MacLeod_2010, Zu_2013}. 
Finally, $\tau_{min}\approx 11$ to a minimal relevant DRW break reported in the work \cite{Smith_2018}. 
The curves are normalized to (-1,1). 
A 2-year-long sample from a mock light curve is presented in Fig. \ref{Figure_interpolation}. 

\section{Results}
\label{Sect_results}
\begin{table}[h!]
  \caption{The median performance in the inference of CDRW spectrum parameters for 1000 different mock light curves with the same $\tau=540, \ \rho=4.77$. The introduced neural inference technique outperforms the baseline MLE in every metric.}
  \label{Table_Inference_performance}
  \centering
  \begin{tabular}{lllllll}
    \toprule
    & \multicolumn{3}{c}{log-Corr. time $log(\tau)$} & \multicolumn{3}{c}{Reverb. time $\rho$}                   \\
    \cmidrule(r){2-4}
    \cmidrule(r){5-7}
    Inference method     & MAD & MAE & RMSE & MAD & MAE & RMSE \\
    \midrule
    Max likelihood & 0.207 & 0.241 & 0.318 & 0.313 & 0.369 & 0.464 \\
    Neural inference & \textbf{0.122} & \textbf{0.157} & \textbf{0.205} & \textbf{0.304} & \textbf{0.354} & \textbf{0.444}  \\
    \bottomrule
  \end{tabular}
\end{table}
\begin{table}[h!]
  \caption{The median performance in the interpolation of CDRW in the same setup as Table \ref{Table_Interpolation_performance}. 
  The mean $\mu(t)$ of interpolation predicted by introduced CDRW and Neural inference is only marginally better than the baseline (MAE, RMSE). 
  However, the uncertainty $\sigma(t)$ is significantly better, which is shown with $\chi^2$.}
  \label{Table_Interpolation_performance}
  \centering
  \begin{tabular}{llllllll}
    \toprule
    & & \multicolumn{3}{c}{Observed periods} & \multicolumn{3}{c}{Gap periods}                   \\
    \cmidrule(r){3-5}
    \cmidrule(r){6-8}
    Model & Inference method & $\chi^2$ & MAE & RMSE & $\chi^2$ & MAE & RMSE \\
    \midrule
    DRW &  Max likelihood & 0.333  & 0.01 & 0.004 & 2.695 & 0.123 & 0.098  \\
    Convolved DRW & Neural inference  & \textbf{0.998} & \textbf{0.008} & \textbf{0.003} & \textbf{0.981} & \textbf{0.121} & \textbf{0.093}       \\
    \bottomrule
  \end{tabular}
\end{table}
\textbf{Neural network.} For training, we simulate a dataset of 300,000 labels $log(\tau)$ and $\rho$ according to eq. \ref{eq_dataset_priors} and split them in train/validation/test in ratio 0.8/0.1/0.1. 
Both stochastic observation times $t^{obs}$ and radiation $F(t^{obs})$ are sampled in the runtime of training to avoid overfitting of random realisations. 
The architecture of the network is presented in Fig. \ref{Fig_Scheme_interpolation_method}. 
In the neural network, the data is interpolated with B-splines on the grid (10,365), then processed by three ResNet blocks \cite{ResNet} with 64,128,256 filters. 
It is followed by linear reduction of the filters to 64 and the attention layer, which is applied by the Learnable Dictionary Encoding (LDE) \citep[]{LDE_1,LDE_2} with 64 clusters.
Finally, the encodings are linearly transformed to predictions of $log(\tau)$ and $\rho$.

The neural network is trained with MSE loss for four epochs until convergence. 
It reaches RMSE 0.15 for $log(\tau)$ and 0.98 for $\rho$ on the test set. 
On Tesla P100, an epoch takes half an hour, given that sampling of the data is done on the GPU. 
We carry out data sampling using Jax \cite{Jax}, the network itself is implemented in PyTorch \cite{PyTorch}.

\textbf{Experiments.} To compare our methods to baseline approaches, we simulate 1000 mock quasar light curves with different stochastic realisations of $t^{obs}$ and $F(t^{obs})$ (eq's. \ref{eq_time_sampling},\ref{eq_dataset_priors}), but the same spectral parameters $\tau$ and $\rho$ for all these curves. 
First, we pick $\tau=540$ as the median correlation time for the observed quasars \cite{Smith_2018}.
Second, we pick $\rho=4.77$, so it results in the break on a month scale, which is midway between DRW $\rho=0$ and the maximal reported break of two months.  

For the first experiment, we compare the performance of Neural Inference (Fig \ref{Fig_Scheme_interpolation_method}) and baseline MLE optimization \citep[TNC;][]{TNC_1,TNC_2} for the inference of $log(\tau)$ and $\rho$. 
The resulting metrics are presented in Table \ref{Table_Inference_performance}. 
In this experiment, Neural Inference outperforms baseline MLE in every metric.

For the second experiment, we compare performance in the interpolation of the same curves on a grid with a regular cadence of one day. 
In this experiment, the baseline approach to quasars is interpolation with DRW, inferred with MLE. 
In turn, the introduced approach is interpolation with CDRW, inferred using the neural inference technique. 
The resulting metrics are presented in Table \ref{Table_Interpolation_performance}. 
The difference between interpolation means $\mu(t)$ (MAE, RMSE) is marginal, but the $\chi^2$ of the new method is significantly better and predicts much more reliable uncertainties than the baseline.

\section{Conclusion}
\label{Sect_conclusion}

In this work, we introduced the new Convolved Damped Random Walk (CDRW) model of quasar radiation (Sec. \ref{Section_CDRW}) that well describes spectra of quasars on all the time scales. 
Additionally, we introduced a new method of inference of Gaussian processes, which we call Neural Inference (Sec. \ref{Section_Neural_inference}). 
This method uses neural networks to mitigate several problems of conventional MLE optimization. 
A combination of these two novelties results in significant improvement in inference and interpolation of a model of a typical quasars light curve (Tables \ref{Table_Inference_performance}, \ref{Table_Interpolation_performance}).


\section{Impact statement}

Impact on Astrophysics:

The new CDRW model decouples the temporal scales of accretion onto a black hole from that of reverberation of the disk surrounding it. 
The introduced reverberation time $\rho$ is the direct tracer of the characteristic radius of the disk $R_{\lambda}$, that traces the mass of the black hole, Eddington luminosity, and accretion efficiency.
Therefore, the CDRW model introduces an explicit constraint on those parameters, which is designed to enhance our understanding of galaxy physics. 

On the other hand, $\rho$ defines the short-scale characteristic time of quasar evolution. 
The distribution of this parameter across quasars can help optimize the observation schedule to maximize the obtained information while minimizing the load on the telescope.

In turn, the Neural Inference approach significantly elevates the quality of inference and interpolation of mock quasar light curves. 
The improvement in inference results in more accurate information about the physics of quasars. 
The improvement in interpolation quality reveals itself in the studies of gravitationally lensed quasars. 
The combination of CDRW and Neural Inference provides better uncertainties that result in the more accurate inference of lensing-induced time delays and, therefore more strict constraints on the Hubble rate.

Impact on other domains:

Time series phenomena permeate our everyday lives -- from finance and healthcare to climate science and astrophysics, since the mathematical formulation of the task is often similar. 
The Gaussian process, especially DRW, is a baseline approach in many fields. 
Moreover, people face the same problems with MLE-based inference in all these fields. 
Therefore the introduced Neural Inference technique can be applied out of the box to constrain the kernel of a Gaussian process for many different types of data.
Additionally, the convolutional term in the CDRW model can be well-adjusted to any data in order to account for domain-specific short-term interaction.

\begin{ack} 



This manuscript has been supported by Fermi Research Alliance, LLC under Contract No. DE-AC02-07CH11359 with the U.S.\ Department of Energy (DOE), Office of Science, Office of High Energy Physics. This research has been partially supported by the High Velocity Artificial Intelligence grant as part of the DOE High Energy Physics Computational HEP program. 

The authors of this paper have committed themselves to performing this work in an equitable, inclusive, and just environment, and we hold ourselves accountable, believing that the best science is contingent on a good research environment.
We acknowledge the Deep Skies Lab as a community of multi-domain experts and collaborators who have facilitated an environment of open discussion, idea-generation, and collaboration. This community was important for the development of this project.

Furthermore, we also thank the anonymous referees who helped improve this manuscript.

\textbf{Author Contributions:} E. Danilov: \textit{Conceptualization, Methodology, Software, Validation, Formal Analysis, Investigation, Data Curation, Writing - Original Draft, Visualization.};
A.~\'Ciprijanovi\'c: \textit{Methodology, Validation, Supervision, Writing - Review and Editing.}; 
B.~Nord: \textit{Conceptualization, Methodology, Investigation, Resources, Writing - Review and Editing, Supervision, Project administration, Funding Acquisition};.
\end{ack}

\newpage
\bibliography{main}

\appendix

\section*{Checklist}

The checklist follows the references.  Please
read the checklist guidelines carefully for information on how to answer these
questions.  For each question, change the default \answerTODO{} to \answerYes{},
\answerNo{}, or \answerNA{}.  You are strongly encouraged to include a {\bf
justification to your answer}, either by referencing the appropriate section of
your paper or providing a brief inline description.  For example:
\begin{itemize}
  \item Did you include the license to the code and datasets? \answerNo{The code and the data are proprietary.}
  \item Did you include the license to the code and datasets? \answerNA{}
\end{itemize}
Please do not modify the questions and only use the provided macros for your
answers.  Note that the Checklist section does not count towards the page
limit.  In your paper, please delete this instructions block and only keep the
Checklist section heading above along with the questions/answers below.

\begin{enumerate}

\item For all authors...
\begin{enumerate}
  \item Do the main claims made in the abstract and introduction accurately reflect the paper's contributions and scope?
    \answerYes{}{We pose problems of the model and the inference \ref{Sect_intro}, propose the solutions \ref{Section_CDRW},\ref{Section_Neural_inference} and test the improvements \ref{Sect_results}. The abstract has the same structure.}
  \item Did you describe the limitations of your work?
    \answerYes{The limitations are given by the distributions used to simulate the data \ref{Sect_data}}
  \item Did you discuss any potential negative societal impacts of your work?
    \answerNo{This is methodology paper for studies of quasars and time series in general. So we don't see any negative societal impacts.}
  \item Have you read the ethics review guidelines and ensured that your paper conforms to them?
    \answerYes{All the used data and papers are public. Human relations are not concerned in the paper.}
\end{enumerate}

\item If you are including theoretical results...
\begin{enumerate}
  \item Did you state the full set of assumptions of all theoretical results?
    \answerYes{The assumptions of quasar light model are given in the cited paper \cite{Sunyaev_QSO_theory}. The mathematical assumptions are given in the section \ref{Sect_data}}
        \item Did you include complete proofs of all theoretical results?
    \answerNo{Convolved DRW spectral form is a direct consequence of the convolution theorem.}
\end{enumerate}

\item If you ran experiments...
\begin{enumerate}
  \item Did you include the code, data, and instructions needed to reproduce the main experimental results (either in the supplemental material or as a URL)?
    \answerYes{We present a public GitHub repository with all the code left from the process of making the paper. It containts all the notebooks needed to reproduce the paper's numbers and figures.}
  \item Did you specify all the training details (e.g., data splits, hyperparameters, how they were chosen)?
    \answerYes{We specify the most important hyperparameters in the section \ref{Sect_results}. Small details are available in the published code.}
        \item Did you report error bars (e.g., with respect to the random seed after running experiments multiple times)?
    \answerNo{}
        \item Did you include the total amount of compute and the type of resources used (e.g., type of GPUs, internal cluster, or cloud provider)?
    \answerYes{In Section \ref{Sect_results}}
\end{enumerate}

\item If you are using existing assets (e.g., code, data, models) or curating/releasing new assets...
\begin{enumerate}
  \item If your work uses existing assets, did you cite the creators?
    \answerYes{}
  \item Did you mention the license of the assets?
    \answerNo{We do not do it explicitly, as all the resources are public.}
  \item Did you include any new assets either in the supplemental material or as a URL?
    \answerNo{}
  \item Did you discuss whether and how consent was obtained from people whose data you're using/curating?
    \answerNA{All the used resources are public.}
  \item Did you discuss whether the data you are using/curating contains personally identifiable information or offensive content?
    \answerNA{Not relevant for study of quasars.}
\end{enumerate}

\item If you used crowdsourcing or conducted research with human subjects...
\begin{enumerate}
  \item Did you include the full text of instructions given to participants and screenshots, if applicable?
    \answerNA{No crowdsourcing}
  \item Did you describe any potential participant risks, with links to Institutional Review Board (IRB) approvals, if applicable?
    \answerNA{}
  \item Did you include the estimated hourly wage paid to participants and the total amount spent on participant compensation?
    \answerNA{}
\end{enumerate}

\end{enumerate}

\end{document}